%% file: driver.tex
\documentclass[%
 reprint,
 amsmath,amssymb,
 aps,
 superscriptaddress
]{revtex4-2}

\newcommand{\vol}{{\ooalign{\hfil$V$\hfil\cr\kern0.08em--\hfil\cr}}}
\newcommand{\eps}{$\langle \epsilon \rangle \,$}
\newcommand{\epstrq}{$\langle \epsilon_{\bar{V}}\rangle$}
\newcommand{\epsB}{$\langle \epsilon_B \rangle\,$}
\newcommand{\logctrd}{$(\, \ln\epsilon - \langle \, \ln\epsilon\,\rangle  ) / \, \sigma(\, \ln \epsilon \, )$}

\usepackage{graphicx}
\usepackage{dcolumn}
\usepackage{bm}

\begin{document}

\preprint{APS/123-QED}

\title{The Excess Dissipation of Energy in a Turbulent Boundary-Layer \\ and its Departure from Log-Normality}

\author{Benjamin Musci}
\email{bmusci3@gatech.edu}
\affiliation{%
 SPEC, CEA, CNRS, Université Paris-Saclay, CEA Paris-Saclay F-91191 Gif-sur-Yvette, France 
 }
\altaffiliation[Supported by ]{Marie-Curie Postdoctoral Fellowship}

\author{S\'ebastien Aumaitre}%
\affiliation{%
 SPEC, CEA, CNRS, Université Paris-Saclay, CEA Paris-Saclay F-91191 Gif-sur-Yvette, France 
 }
 
\author{Enzo Francisco}
\affiliation{%
 SPEC, CEA, CNRS, Université Paris-Saclay, CEA Paris-Saclay F-91191 Gif-sur-Yvette, France 
 }

\author{Adam Cheminet}
\affiliation{%
 SPEC, CEA, CNRS, Université Paris-Saclay, CEA Paris-Saclay F-91191 Gif-sur-Yvette, France 
 }%
 
\author{B\'ereng\`ere Dubrulle}
 \email{berengere.dubrulle@cea.fr}
\affiliation{%
 SPEC, CEA, CNRS, Université Paris-Saclay, CEA Paris-Saclay F-91191 Gif-sur-Yvette, France 
 }%


%



\date{\today}

\begin{abstract}
We investigate turbulent dissipation in a von Kármán flow using PIV and Diffusing Wave Spectroscopy measurements to directly compare bulk and wall dynamics. 
While bulk dissipation conforms to the dissipative anomaly, wall dissipation exhibits a clear excess that grows with $Re$, consistent with velocity-gradient–dominated scaling.  Decomposition into dissipation intensity bands reveals that this excess is mainly driven by progressive redistribution toward high-intensity events ($>$10 \eps⟩) as $Re$ increases.
From these measurements, we infer the skin-friction coefficient, finding a decreasing trend with $Re$ fairly consistent with classical power-law behavior despite increasing dissipation.
Statistically, the wall shows strong departures from log-normality at low $Re$ that diminishes with increasing $Re$, reflecting an increase in the effective dimensionality of the near-wall gradient field with $Re$. In contrast, the bulk dissipation remains near log-normal across all $Re$ with slowly growing log-dissipation variance, consistent with K62 refined similarity. 
These results suggest distinct origins of log-normal behavior: multiplicative cascade dynamics in the bulk versus the combined effect of persistent shear and a superposition of an increasing number of independent gradient contributions at the wall.

\end{abstract}

\maketitle


\input{sections/intro}

\input{sections/exp_facility}

\input{sections/discussion_v2}


\begin{acknowledgments}
The authors thank  C\'{e}cile Wiertel-Gasquet for her countless efforts in the coordination and execution of the experiments, as well as her patience with the first author. They also thank Vincent Padilla for his practical wizardry and quick problem solving. Thanks are especially due to Murukesh Muralidhar for his help and flexibility concerning the use of the high-speed camera. Thanks is given to the French government grant managed by Agence Nationale de la Recherche France 2030 under the reference ANR-24-RRII-0001, as well as the ANR BANG grant agreement no.ANR-22-CE30-0025. Finally the first author is thankful for the generous support of the postdoctoral fellowship through the European Union’s Horizon 2020 research and innovation program under the Marie Sklodowska-Curie grant agreement No. 945298.
\end{acknowledgments}

\bibliography{dws_prl,SM}

\end{document}

%% file: sections/intro.tex
\paragraph*{Introduction} The spatiotemporal nature of energy dissipation in real turbulent flows contains several open questions. 
Chief among them is the universality of dissipation mechanisms and statistics compared to those in homogeneous isotropic turbulence (HIT) \cite{Meneveau1991, Jimenez2013}.
The gradient-squared nature of the turbulent kinetic-energy dissipation rate, $\epsilon$, means that small changes in the distribution tails of turbulent velocity gradients can have an out-sized impact on dissipation behavior, which is further compounded by flow anisotropy or boundaries. 

The dissipation of turbulent energy at walls is of particular interest, as all real turbulent flows are bounded in some way. 
Elucidating whether the wall region has its own dissipation statistics and scaling laws (PDF shape, intermittency measures), or whether it smoothly connects to bulk-flow/HIT behavior after the right normalization, is thus a crucial question \cite{Jimenez2013, Benzi1999, Duan2025}. 
Strong wall shear and geometric constraints violate HIT assumptions of local universality \cite{Kolmogorov1941} (hereafter K41). Despite vanishing velocity at the wall, dissipation remains finite and fluctuating, yet wall effects on intermittency remain poorly understood \cite{Duan2025}.

\begin{figure*}
\centerline{\includegraphics[scale=0.65]{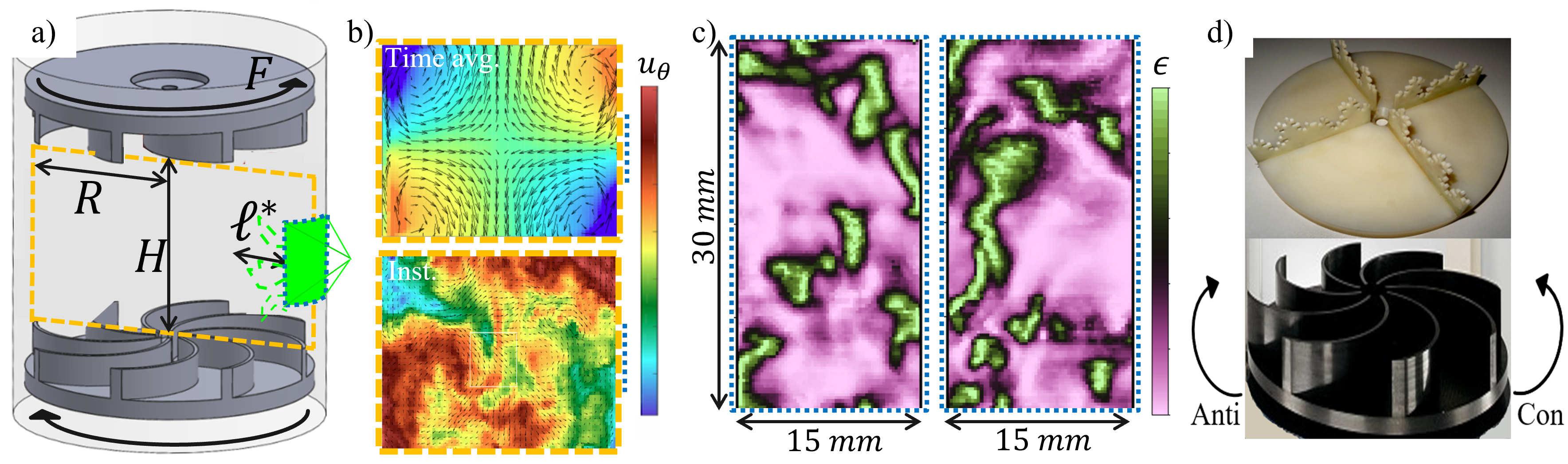}}
\caption{\label{fig:facility} (a) Schematic of experimental facility (not to scale): $H$=0.18m, $R$=0.1m, $\ell^*$=80$\mu$m. (b) Time-averaged and instantaneous meridional velocity fields. Arrows denote in-plane velocity and color denotes azimuthal velocity $u_\theta$. (c) Examples of $\epsilon(\theta,z,t)$ provided by DWS at two different $Re$. (d) Impellers used in this work: top is ``Fractal"  while bottom is the direction-dependent ``Anti"/``Con".} 
\end{figure*}

From a modeling perspective, the rate of dissipation is directly linked to drag/power loss, thermal wall loading, mixing intensity, and peak shear fluctuations crucial for fatigue and vibration considerations \cite{Delafosse2011, Ahn2010, Bose2018, Kakka2020, Moser2021}.
For instance, Large Eddy Simulation models that rely on sub-grid-scale (SGS) closures simply extending bulk-turbulence behavior to walls will fail to capture departures from universality such as the bursting, enhanced instability, and non-local interactions observed near walls \cite{Sarie1994, Benzi1999, Jimenez2013, Agostini2016, Bose2018, Lund1994}.
Measuring the dissipation directly and quantifying how its statistics differ between the wall and bulk therefore provides direct constraints for SGS and wall-stress closures, especially concerning extreme events and Reynolds-number trends that integrated-dissipation metrics cannot reveal \cite{Hack2019, Moser2021, Kakka2020, Schroder2024}.

Fundamentally, the nature of dissipation in the inviscid limit of turbulence is an open question, as it depends crucially on the flow geometry \cite{Drivas2018, Eyink2024}. 
Flows over rough surfaces obey the so-called ``0th-law of turbulence,'' in so far as they continue to dissipate energy, on average, at a finite rate as the Reynolds number ($Re$) increases despite the simultaneous decrease in viscous effects \cite{Iyer2025, Eyink2024}. 
In contrast, flows over smooth surfaces have been characterized by skin friction ($C_f$) laws decreasing with $Re$ \cite{McKeon2005, Dubrulle2024, Eyink2024}, while periodic flows display a very weak variation with $Re$ that may or may not converge to 0th-law behavior \cite{Iyer2025}. 
Whether this can be resolved via the non-smoothness of the velocity field in the bulk, or by way of rapidly concentrated velocity gradients in a vanishingly small layer along flow boundaries, remains unknown \cite{Kato1984, Kelliher2007, Drivas2018, Iyer2025}. 
Thus, an open question is whether dissipation is primarily a bulk phenomenon driven by intermittent structures or becomes localized in boundary-attached layers without requiring the velocity-field roughness hypothesized for the bulk by \citet{Onsager1949}.

The review by \citet{Eyink2024} identifies this as a gap in turbulence theory: existing frameworks link bulk anomalous dissipation to limited velocity-field regularity, but provide no analogous closure for walls, where studies have observed differing behavior across boundary conditions and roughness's.
Currently there is insufficient evidence, numeric or experimental, to settle whether a wall dissipation anomaly exists or how it might scale with $Re$. 
This is because direct experimental access to the spatiotemporal dissipation at a boundary has remained prohibitively challenging, especially in conjunction with a clean bulk comparison in the same flow \cite{Delafosse2011, Francisco2023, Schroder2024, Duan2025, Kuzzay2015}.

This Letter seeks to address these questions by applying a novel diagnostic, Diffusing-Wave Spectroscopy (DWS), at the boundary of a von K\'{a}rm\'{a}n flow. 
The DWS technique provides unparalleled access to spatiotemporal dissipation at flow boundaries compared to previous experimental approaches, and we use it to directly compare dissipation statistics and distributions between the wall and flow bulk.

%% file: sections/exp_facility.tex
\begin{figure*}
\centerline{\includegraphics[scale=0.5]{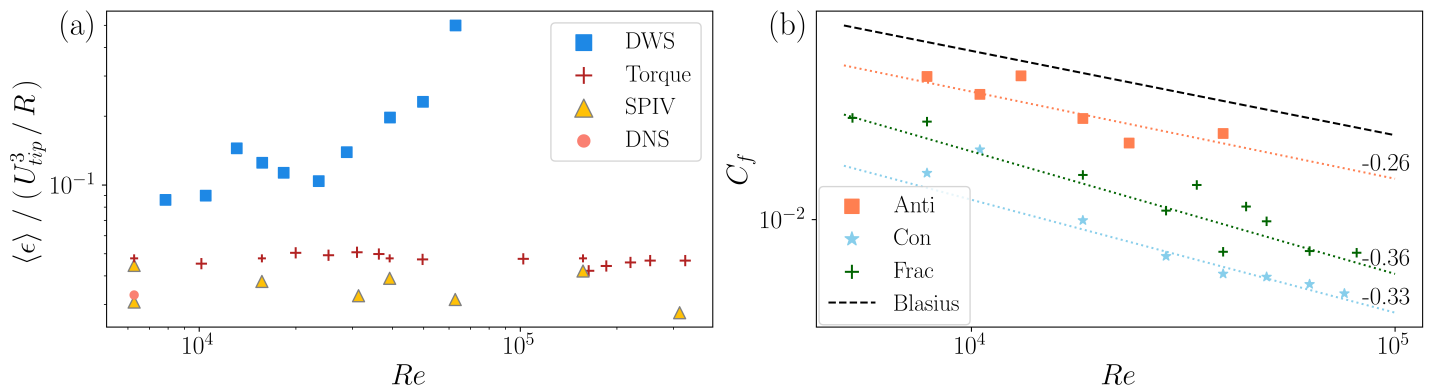}}
\caption{\label{fig:mean_eps} a) Measured values of $\langle \epsilon \rangle$ relative to the theoretical HIT dissipation value, $\epsilon_{HIT}$ predicted by K41. Values are ensemble averages at each $Re$. Blue squares: wall DWS \eps; yellow triangles: bulk TPIV \epsB; red pluses: volumetric torque-meter dissipation \epstrq. 
b) Skin friction coefficient measured by DWS for 3 different impeller types: Anti, Con, and Frac \cite{supp}. Power law fits are shown as dotted lines, with the exponent shown at the end. The classical Blasius relation is the blacked dashed line.  }
\end{figure*}

\paragraph*{Experiment and Diagnostic} We use the von Kármán flow experiment at CEA Paris-Saclay to obtain the experimental data presented here. 
The facility has been used extensively to study turbulence and extreme event statistics at the flow center \cite{Ravelet2004, Kuzzay2015, Saw2016, Saint-Michel2013,Debue2018, Saw2018}. 
The cylindrical facility (Figure \ref{fig:facility}a) contains constant temperature ($20^{\circ}C$) water as the working fluid, and uses two counter-rotating impellers  to drive the flow to steady-state turbulence \cite{Ravelet2004}.
Two curved blade (termed ``Anti" and ``Con") and one straight blade (``Frac" for Fractal) impellers are used in this work (see Supplemental Material \cite{supp}). 
The tank has a radius ($R$) and height (distance between impellers) of 0.1m and 0.18m, respectively.
The integral scale Reynolds number ($Re = 2\pi R^2 F / \nu$) is changed via the impeller rotation frequency, $F$.  The resulting flow is fully turbulent with a coherent, large-scale structure in the time-averaged sense (Figure \ref{fig:facility}b) which contains a homogenized and quasi-isotropic shear layer in the mid-plane of the cylinder. 
At the center of this shear-layer, previous work using Tomographic Particle Image (TPIV) and Particle Tracking Velocimetry (PTV) have allowed for the computation of higher-order turbulent statistics shown to agree well with HIT DNS results \cite{Geneste2019, Debue2018, Debue2021,Musci2025}.
The global bulk dissipation \epstrq is obtained from torque-meters on the two impellers \cite{Marie2004, Kuzzay2015, Dubrulle2019, Dubrulle2022}. 

Bulk dissipation is estimated from the TPIV data using a scale-resolved formulation of the weak Kármán–Howarth–Monin equation, $\epsilon_B \approx \mathcal{D}^I_\ell + \mathcal{D}^\nu_\ell$, which locally separates inertial and viscous transfer without direct differentiation of the velocity field \cite{Dubrulle2019, supp}.

Crucially DWS allows direct, time-resolved measurement of the norm of the strain-rate tensor at flow boundaries.
For Newtonian fluids, this can be related to the viscous dissipation rate of energy, $\epsilon$, via $\epsilon = 2 \nu \sum_{i,j} \, S_{ij}^2 $.

Here $\nu$ is the kinematic viscosity and $\, S_{ij} = \frac{1}{2} (\partial u_i / \partial x_j + \partial u_j / \partial x_i)$ is the strain rate tensor, $  \nabla \bm{u} = \partial u_i / \partial x_j $. 

DWS infers local dissipation from the temporal auto-correlation of multiply scattered light, which encodes characteristic Brownian and turbulence-driven time scales. 
Experimental validity requires tracer particles in the ballistic regime (low Stokes number) and a scattering length scale $\ell^*$ smaller than the smallest relevant flow scales, ensuring the measurement samples the turbulent boundary layer. 

Full details of the methodology, calibration of the scattering depth, and DWS validation criteria for the range of $Re$ explored in this study (6,000 - 80,000) are provided in the Supplemental Material \cite{supp}.

%% file: sections/discussion_v2.tex
\bigskip
\paragraph*{Results and Discussion}

\paragraph{Coherent Dissipation Structures} ~\\
The dissipation fields in Figure~\ref{fig:facility}c show that dissipative action at the wall occurs within coherent structures of preferred directionality and shape.
The structures are typically stretched along an axis that aligns most with the vertical axis of the facility - perpendicular to the azimuthal driving direction of the impellers.
Further, as expected, the size of these structures generally decreases with $Re$. 
The presence of coherent structures is notable as research by \citet{Dubrulle2019} highlighted the heightened importance of local dynamics and indicated that some of the highest energy transfer and dissipative events occur around coherent structures. 
Conversely, regions outside coherent structures resulted in behavior close to K41's approximation with reduced intermittency \cite{Saint-Michel2014, Dubrulle2019}.

\bigskip
\paragraph{Excess Dissipation and $Re$ Trend}~\\
Concerning dissipation statistics, we first consider the spatial-temporal average of the kinetic energy dissipation rate, \eps. 
The spatial extent of the averaging is indicated in Figure~\ref{fig:facility}a-c by the dotted blue line/rectangle. 
The DWS wall measurements are shown in Figure~\ref{fig:mean_eps}a normalized by the theoretical dissipation rate for HIT flow: $\epsilon_{HIT} \propto U^3_{tip}/R$, where $U_{tip}$ is the velocity of the impeller tip at a given $Re$ \cite{Kolmogorov1941,Saint-Michel2014}.
The dissipation rate measured by the facility's torque-meters, \epstrq, is also shown in Figure~\ref{fig:mean_eps}a, along with the dissipation rate at the center of the von K\'{a}rm\'{a}n flow, \epsB, obtained using SPIV and TPIV diagnostics in previous works \cite{Debue2018, Debue2021, Geneste2019, Musci2025}.

As observed in previous studies \cite{Dubrulle2019, Dubrulle2022, Kuzzay2015, Saw2018}, when normalized, \epstrq and \epsB approach a constant for $Re > 6000$; a manifestation of the dissipative anomaly \cite{Iyer2025, Dubrulle2019, Drivas2018, Yao2020, Eyink2006}.
Of the three dissipation measurements shown, \eps at the wall is always the largest, and thus in reference to the average volumetric values \epstrq, we call this an ``excess'' of dissipation at the wall. 
While it has been hypothesized that high shear at the wall should locally increase turbulent kinetic energy production that must be dissipated, it is not obvious that this should occur at the wall itself, yet these measurements show that it does. 

\begin{figure*}
\centerline{\includegraphics[scale=0.49]{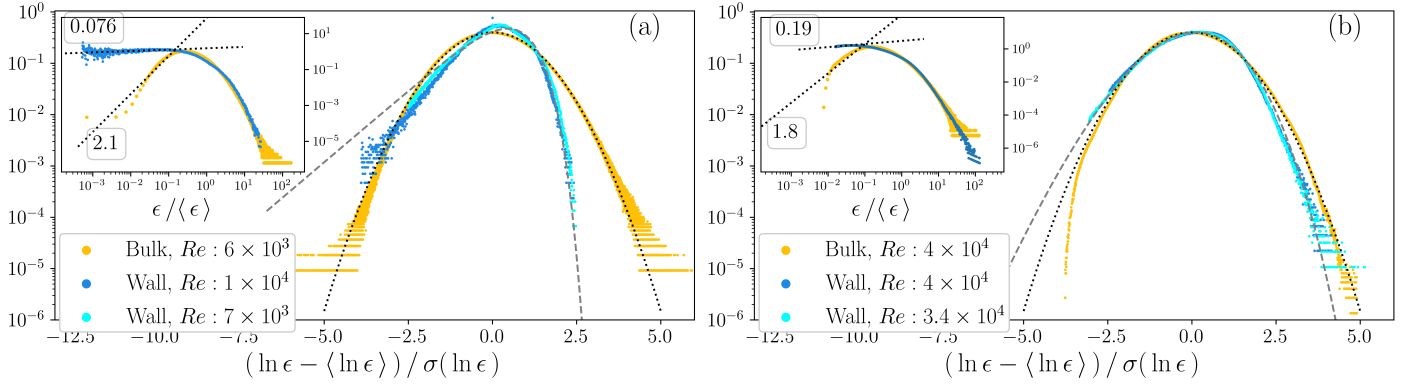}}
\caption{\label{fig:PDFs} PDFs of normalized dissipation, $\epsilon / \langle \epsilon \rangle $ (insets), and centered-reduced dissipation, \logctrd, at two $Re$ ( a) 6000, b) 40000) for the DWS wall measurements (blue) and TPIV bulk measurements (gold). Dotted black line shows a log-normal distribution in the main plots, while in the insets it is a power-law fit to the left-tail, with the exponent displayed. Dashed-black lines show a Chi-squared ($\chi_k^2$) distribution with 2 and 30 terms for a) and b), respectively. 
}
\end{figure*}

Furthermore, \eps is clearly increasing with $Re$, meaning the excess dissipation grows in the same flow where we observe 0th-law behavior in the volumetric and bulk measurements, \epstrq and \epsB. 
The contemporary DNS work of \citet{Duan2025} found qualitatively similar results, observing near 0th-law behavior in the outer boundary layer region while finding excess dissipation at the wall that grew with $Re_\tau$ according to a ``bounded defect law.''
This stands in apparent contrast to the DWS measurements of \citet{Francisco2025}, who reported boundary dissipation scaling with injected power in a cube geometry. 
However, their measurement location near the impeller likely compresses the boundary layer substantially relative to our measurements at the tank center, such that their $l^*$ may include the viscous sublayer and the outer cascade-dominated region, allowing for an effectively bulk-like scaling.

 It is noteworthy that we observe finite and increasing dissipation at a smooth wall as $Re$ grows, as some works have observed decreasing dissipation for smooth walls, with others reporting 0th-law behavior for rough walls, wakes, and various boundary conditions \cite{Eyink2024}. 
More importantly, walls introduce new mechanisms that can generate or suppress dissipation (via additional energy flux terms at the boundary) independent of any bulk behavior.
The trends we observe are thus indicative of additional dissipation pathways and highlight a gap in current turbulence theory since we simultaneously observe anomalous dissipation in the bulk - a mechanism linked to limited regularity of the velocity field.

In the least, these results suggest that as $Re$ increases, boundary layer dissipation becomes 
increasingly decoupled from the integral-scale velocity $U$. 
We hypothesize that velocity gradients at the wall steepen continually with $Re$, so that wall dissipation becomes increasingly gradient-dominated and distinct from the bulk-cascade. 
Taking a dimensional-analysis perspective, we assume:
\begin{equation}
    \epsilon_{wall} \approx u_\tau^4 / \nu \;,
    \label{eq:eps_dim}
\end{equation}
\noindent where $u_\tau = \sqrt{\tau_\omega / \rho}$ is the friction velocity. 
Linking $u_\tau$ to $U$ via the skin friction coefficient  $C_f = 8\tau_\omega / \rho U^2$, and using the classical result $C_f \propto Re^{-n}$ with $n < 0.25$ \cite{White2006, Schlichting2016}, one 
can estimate $\epsilon_{wall} \approx U^3 Re^{1-2n}/R$. 
Then, as long as $n < 0.5$, we can expect $\epsilon_{wall}/\epsilon_{HIT}$ to grow with $Re$, consistent 
with Figure~\ref{fig:mean_eps}a.

This line of reasoning also shows that DWS allows for experimental estimation of $C_f$. 
From Equation~\ref{eq:eps_dim} and the relations above, $C_f \approx 8\sqrt{\langle\epsilon\rangle R \, / \, U^3 Re}$, a decreasing trend which is observed in Figure~\ref{fig:mean_eps}b for all forcing/impeller configurations.
As empirical relations for $C_f$ do not exist for the von K\'{a}rm\'{a}n flow, we use the Blasius power-law $C_f = 0.3164\,Re^{-1/4}$ for an illustrative comparison \cite{Blasius1913, McKeon2005, Dubrulle2024}. 
The power-law fits to the data in Figure~\ref{fig:mean_eps}b show $C_f$ behavior in reasonable agreement with this trend.
This behavior further establishes DWS as a viable technique for directly estimating $C_f$.

The different absolute values of $C_f$ for each forcing at a given $Re$ can be equated to an effective boundary roughness caused by the different impellers.
Previous works have found Anti impellers to produce higher turbulent intensities than Contra for the same integral $Re$ \cite{Ravelet2004, Saint-Michel2013, Dubrulle2022}. 
This then results in different volumetric dissipation and thus different Kolmogorov and Taylor length scales at a given $Re$.
Thus, $C_f$ is linked more intimately to turbulent fluctuations than to integral-scale forcing alone.

Furthermore, the observation that $C_f$ decreases with $Re$ while dissipation increases is non-trivial: velocity gradients are amplified sufficiently by turbulence to increase dissipation despite a simultaneous weakening of wall stress relative to $U^2$. 
Finally, these results add nuance to the findings of \citet{Cadot1997}: the coexistence of decreasing $C_f$ and increasing wall dissipation arises because wall dissipation is controlled by velocity gradients tied to skin friction rather than by $U$ directly, such that both trends can naturally coexist.

\bigskip
\paragraph{Distributions and higher order moments of $\epsilon$}~\\
For further insights into the dissipation at the wall and bulk we show the normalized ($\epsilon/$\eps) and center-reduced (\logctrd) dissipation distributions at low and high $Re$ in Figure~\ref{fig:PDFs}. 
We first note that the normalized distributions (inserts) show greatly varying left tail characteristics for the wall and bulk. 
Generally, at the wall the left tail slope is near 0, indicating a large influence of near-zero dissipation events. 
As $Re$ increases (Figure~\ref{fig:PDFs}b), this slope grows and the dominance of near-zero events decreases. 
Conversely, near the flow center the steeply positive left tail slopes indicate a vanishing PDF for zero dissipation events, meaning the bulk is more strongly protected against near-zero dissipation. 
These differences may hint at varying mechanisms of dissipation organization for the wall versus the bulk, where the high shear rates at the wall may serve to counterintuitively enhance the probability of low-dissipation events relative to the bulk.

Previous works have found similar results for the bulk, arguing that HIT flow should have a left tail power-law exponent near 1.5 \cite{Duan2025, Gotoh2023, Elsinga2023}. 
Departure from this value is argued to reflect the relative number of velocity-gradient terms actively contributing to dissipation, where a slope near 0 corresponds to two contributing gradient terms at the wall.
Likewise, a slope of 2 for the bulk indicates contribution from six terms \cite{Gotoh2023}. 
This suggests the wall/bulk left-tail difference should  be read as a difference in the kinematic structure of the dissipation field: since $\epsilon \approx 0$ requires all contributing velocity gradients to be simultaneously negligible, this becomes increasingly improbable as more gradient terms are active. 
Thus, any non-zero dissipation at the wall is likely built from only a few independent velocity gradients, allowing for both the diverging left tail and the excess \eps seen in Figure~\ref{fig:mean_eps}a.

Concerning the center-reduced distributions in the main panels of Figure~\ref{fig:PDFs}, the bulk PDFs remain close to log-normal (dotted line) for both $Re$ cases, consistent with \citet{Kolmogorov1962} (K62). 
The wall, by contrast, displays strong asymmetry at low $Re$ (Figure~\ref{fig:PDFs}a) with a prominent left tail, a relatively weak right tail, and a narrow, rightward-shifted peak; a distinct departure from log-normality.
The distribution is more closely described by a ``Chi-squared" ($\chi^2_k$) distribution with $k=2$ (dashed line), whose right tail is particularly well captured. 
This directly supports the ``two-contributing-terms'' interpretation of \citet{Gotoh2023, Ruiz-chavarria2000}: both the low- and high-dissipation events are governed by two dominant, near-Gaussian velocity-gradient components, per the $\chi^2_k$ definition.
However, the left tail decays somewhat faster than the $\chi^2(k=2)$ prediction and the peak is higher and narrower (readily apparent in a linear scale, although not shown here), indicating this is not a wholly adequate model. 
The distribution instead suggests there is a persistent background dissipation level (due to sustained shear or organized boundary layer) on which fluctuations are superimposed, rather than a purely stochastic low-dimensional process.

At higher $Re$ (Figure~\ref{fig:PDFs}b), the degree of asymmetry at the wall has clearly lessened (though it remains more pronounced than in the bulk), and the distribution's peak aligns more closely with both the log-normal and $\chi^2_k$ distributions, signifying the erosion of a preferred dissipation level.
Both tails of the wall dissipation are now more closely described by a $\chi^2_k$ distribution with $k \approx 30$. 
One should not interpret this as 30 velocity-gradient terms contributing to dissipation, as resolution, coarse-graining, averaging, etc.\ could be at play at the higher $Re$ case.
However, one can confidently say the number of degrees of freedom in the wall dissipation has increased, where \eps is now a more composite quantity at higher $Re$.
This may reflect more independent gradient contributions, additional small-scale structures, or increased cascade-like multiplicativity.
Additionally, the increasing log-normality of our wall PDFs at higher $Re$ is consistent with the $\ell^*$-related interpretation of the  \citet{Francisco2025} disagreement discussed above, and elaborated below.

\begin{figure}
\centerline{\includegraphics[scale=0.44]{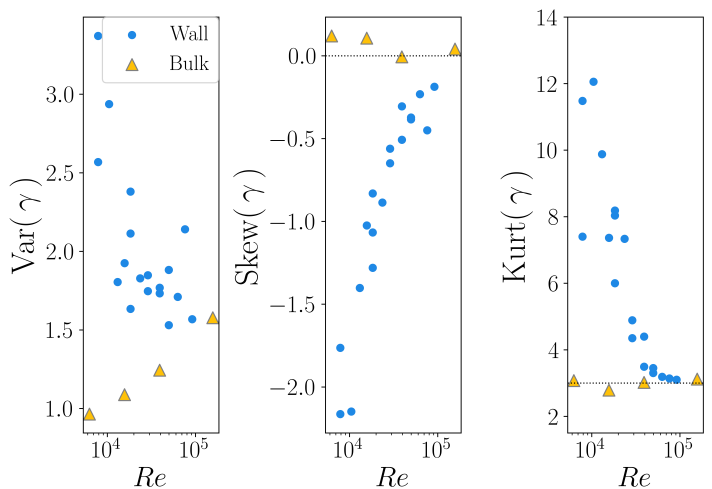}}
\caption{\label{fig:stat_moms} 2nd-4th order statistical moments of $\gamma = \ln ( \epsilon / \langle \epsilon \rangle )$ for wall DWS (blue circles) and bulk PIV (gold triangles) measurements.  }
\end{figure}

To better interpret the different trends of mean dissipation in Figure~\ref{fig:mean_eps}, we also examine the variance, skewness, and kurtosis of $\gamma = \ln(\epsilon/\langle\epsilon\rangle)$ for the wall and bulk, shown in Figure~\ref{fig:stat_moms}. 
At the wall, there are trends of decreasing variance, increasing skewness, and decreasing kurtosis with $Re$, whereas in the bulk, the variance increases slowly while skewness and kurtosis remain essentially constant.

For the bulk, the near-constant skewness near 0 and kurtosis near 3 alongside slowly growing variance are consistent with a distribution remaining close to log-normal and broadening gradually with $Re$, in line with K62. 
At the wall, the initially large variance reflects broad excursions in log-dissipation at low $Re$; the strongly negative skewness indicates a distribution pulled heavily toward low-dissipation events, with a sharp, narrow peak offset to the right.
As $Re$ increases, variance decreases, skewness approaches zero, and kurtosis falls toward 3.
Thus the $\gamma$ distribution becomes more symmetric and less dominated by low-dissipation events, approaching a more Gaussian shape in log-space, consistent with Figure \ref{fig:PDFs}b.

\begin{figure}
\centerline{\includegraphics[scale=0.59]{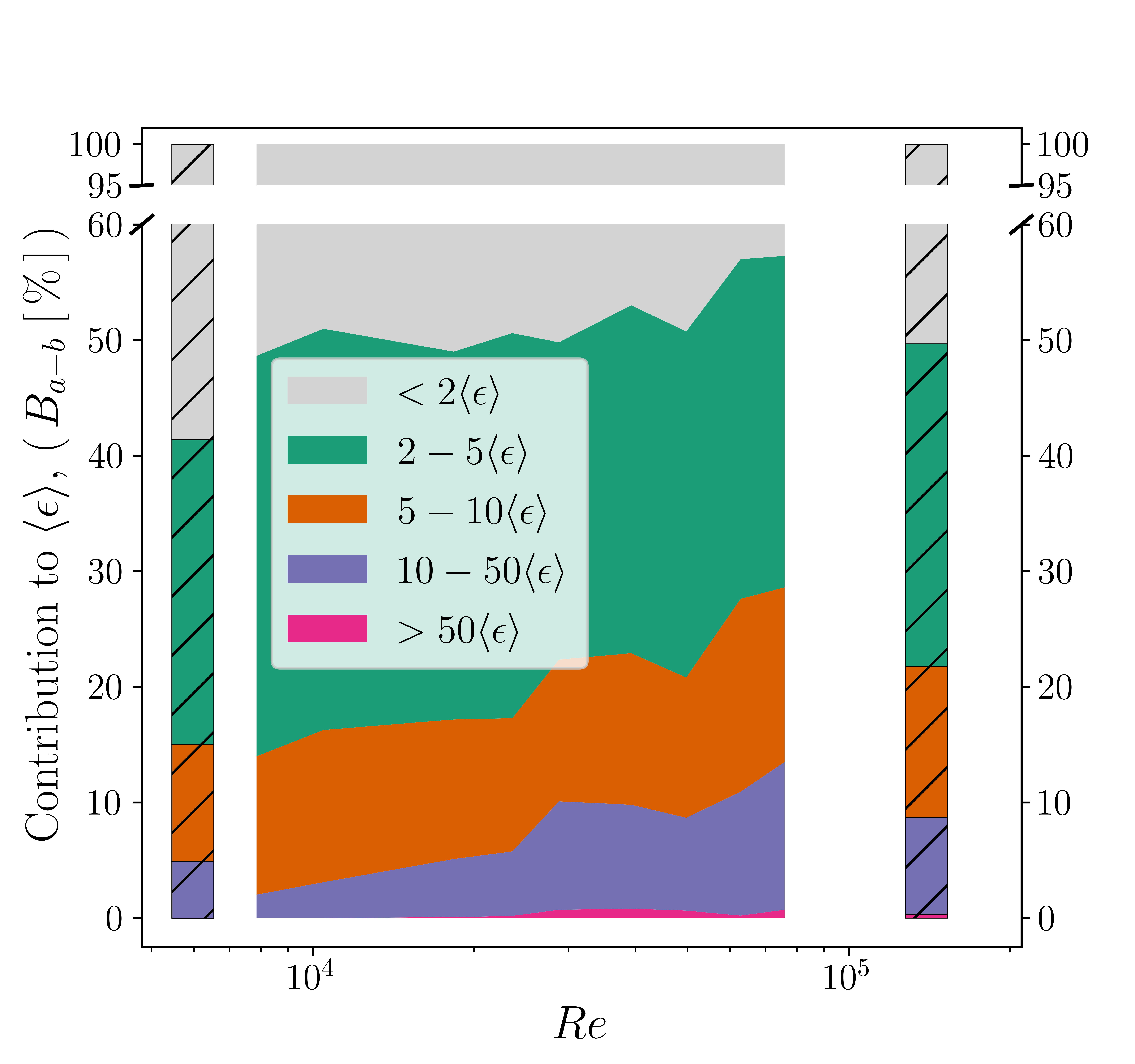}}
\caption{\label{fig:ContBnds} 
Contribution of intensity-banded events to the mean \eps shown in Figure \ref{fig:mean_eps} for the bulk (hatched bars) and the wall (solid bars). The bulk bands undergo much smaller changes across the probed $Re$ range and are thus only shown at the two limits of their range. 
}
\end{figure}

\bigskip
\paragraph{Interpretation of Excess Dissipation}~\\
Overall, the $\gamma$ moment trends at the wall display a change from highly non-log-normal distributions toward more Gaussian ones at higher $Re$. 
A naive reading might interpret this as decreasing wall intermittency with $Re$. 
However, intermittency is properly defined through deviations of velocity increments from Gaussianity, not through dissipation PDF shape, and the results of Figure~\ref{fig:ContBnds} provides a concrete reason to resist this interpretation.
Figure~\ref{fig:ContBnds} quantifies to what extent events within a dissipation band contribute to the mean; that is $B_{a-b} = \sum_{a < z_i < b} \epsilon_i / \sum_i \epsilon_i$, where $z_i = \epsilon_i / \langle \epsilon \rangle$.
The band contributions for the bulk (hatched bars) are largely invariant across the probed $Re$ range, reflecting the self-similar cascade statistics already evidenced by the near-constant $\gamma$ moments in Figure \ref{fig:stat_moms}. 
The wall, by contrast, undergoes a clear redistribution: at low $Re$, wall dissipation is dominated by events $< 2\langle\epsilon\rangle$ and between $2$-$5\langle\epsilon\rangle$ ($B_{<2} + B_{2-5}$ $\approx 85\%$). 
As $Re$ increases, these lower-intensity contributions shrink to $\approx 70\%$, displaced by stronger dissipative events $> 10\langle\epsilon\rangle$, with $B_{10\text{-}50}$ alone growing from a $2\%$ to $13\%$ contribution. 
Thus the rising \eps observed in Figure \ref{fig:mean_eps}a can be linked to increased occurrence of intense events, which could be interpreted as intermittency since localized high dissipation events contribute more disproportionally to \eps at the wall.  
Note that to the authors' knowledge, the decomposition of \eps into banded intensity mean-contributions ($B_{a-b}$) has not been previously reported, though related threshold-based analyses exist across the literature \citep{Yeung2015, Elsinga2023}.

That the distribution simultaneously becomes more log-normal is not in conflict with this.
At low $Re$ the wall right tail decays faster than log-normal, so the transition toward log-normal statistics necessarily entails a heavier right tail and more extreme events. 
The apparent tension between increasing log-normality and increasing extreme-event contributions is precisely what demonstrates that PDF departure from log-normality is a poor proxy for intermittency as, at the wall, the two evolve in opposite directions with $Re$.

Rather than reflecting changes in intermittency, the evolution of the $\gamma$ shape parameters in Figure \ref{fig:stat_moms} is more plausibly explained by an increase in the effective dimensionality of the near-wall gradient field. 
The transition from $\chi^2(k=2)$ to $\chi^2(k=30)$ observed in Figure~\ref{fig:PDFs} directly reflects this: at low $Re$, only a small number of independent gradient contributions dominate the dissipation field, producing the strongly asymmetric, low-dimensional statistics observed. 
As $Re$ increases, more independent gradient contributions combine to produce dissipation, and this superposition drives the distribution toward log-normality regardless of whether a cascade mechanism is present. 
This is consistent with the $C_f$ trends of Figure \ref{fig:mean_eps}b and the findings of \citet{McKeon2005} and \citet{Dubrulle2024}, which emphasizes that skin friction increasingly depends on the specific velocity profile structure as $Re$ increases, suggesting more structures contribute to the near-wall velocity gradient at higher $Re$. 
Thus, wall log-normality at high $Re$ likely reflects structural superposition at the boundary rather than any mechanism specific to the bulk's inertial cascade. 
Crucially, this decouples the shape evolution of the log-dissipation distribution from the behavior of extreme events: while the $\gamma$ shape parameters indicate a more symmetric and regular distribution at high $Re$, the band decomposition shows that extreme dissipation events ($>10\langle\epsilon\rangle$) contribute increasingly to the mean, and the PDF right tail extends further in standardized space. 
If anything, the evidence here points toward intensifying extreme-event activity at the wall with $Re$, even as the overall distribution shape normalizes.

It should however be noted that for a fixed $\ell^*$ DWS campaign such as this one, the boundary layer thins with increasing $Re$ while the optical probing depth $\ell^*$ remains constant, meaning that the ratio of the measurement volume to the viscous sublayer grows with $Re$. 
While we estimate that $\ell^*$ remains smaller than the viscous sublayer for the range of explored $Re$ (see Supplemental Material \cite{supp} Table 1), these are only estimates and can not be confirmed with the present dataset.
The observed trend toward log-normality at the wall may therefore reflect, at least in part, increasing bulk contamination of the DWS signal rather than purely structural evolution of the near-wall dissipation field; a degeneracy that cannot be fully evaluated at present.

The relationship between PDF shape and intermittency at the wall is therefore not straightforward, and our results caution against conflating the two. 
Work by \citet{Elsinga2023} suggests that higher-order moments in the bulk are primarily controlled by sparse but intense large-scale shear layers, sitting in a quieter dissipation background.
However at the wall,  shear layers are a permanent boundary condition rather than an intermittent mechanism; they are persistent and space-filling rather than sparse and intense. 
This could explain how the wall simultaneously sustains a strongly distorted dissipation PDF and elevated mean dissipation, while the departure from log-normality reflects persistent shear-driven spatial organization rather than intermittency in the cascade sense. 
This interpretation is consistent with the findings of \citet{Duan2025}, who showed the wall region to be relatively non-intermittent, and with the emphasis of \citet{Dubrulle2019} on local inertial energy transfer as the more appropriate lens through which to understand intermittency corrections. 
Near the wall, persistent geometry-imposed shear strongly reshapes the dissipation field, breaks HIT/local universality, and changes the distribution shape, yet may not automatically produce stronger high-order intermittency.

In summary, these results show that near-wall dissipation exists as an excess relative to the bulk which increases with $Re$. 
Wall dissipation PDFs evolve from low-degree $\chi^2$ (Rayleigh) statistics governed by a few dominant gradient components toward higher-degree $\chi^2$ behavior reflecting a superposition of many contributions. 
This points toward dissipation generated by distributed boundary-layer gradients rather than being tied to the same regularity criteria that govern the bulk anomaly. 
It further suggests that log-normality and intermittency need not be strongly coupled in wall-bounded flows, and that intermittency corrections may be better understood through coherent structures and local dynamics.
Assuming the validity of Equation~\ref{eq:eps_dim}, our findings further suggest that a unified description of wall turbulence must simultaneously reconcile $C_f \propto Re^{-n}$ and $\epsilon \propto Re^{1-2n}$ with this evolving statistical organization, a description not yet provided by existing theory \citep{Eyink2024}.